\documentclass[amsmath,amssymb]{natureprintstyle}
\usepackage{graphicx}
\usepackage{ulem}
\usepackage{amsmath}
\usepackage{amssymb}
\usepackage{color}
\usepackage{colortbl}
\usepackage{subfig}
\usepackage{mathtools}
\usepackage[T1]{fontenc}
\usepackage[latin9]{inputenc}
\usepackage{babel}
\usepackage[colorlinks, citecolor=blue]{hyperref}
\usepackage{comment}

\usepackage{mdframed}
\usepackage[svgnames]{xcolor}
\definecolor{myframecolor}{HTML}{DAF7A6} 
\usepackage{multicol}

\title{%Rota interior : 
The Hidden Wheel-Within % the Material
}

\author{Falko Ziebert $^{1,2}$ and Igor M. Kuli\'{c} $^{3*}$}

\begin{document}

\maketitle 

\begin{affiliations} 
	\item Institute for Theoretical Physics, Heidelberg University, Philosophenweg 19, 69120 Heidelberg, Germany
	\item BioQuant, Heidelberg University, Im Neuenheimer Feld 267, 69120 Heidelberg, Germany
	\item Institut Charles Sadron UPR22-CNRS, 67034 Strasbourg, France
	*Correspondence should be addressed to kulic@unistra.fr
\end{affiliations}

{\bf 
There is this old, eternal %ever reoccurring 
question: Why don't animals
have wheels? In this perspective we show that
they actually do. And they do so in a physically extraordinary way
-- by combining  incompatible elasticity, differential geometry and dissipative self-organization. 
Nature's %favorite 
wheel -- the ``wheel-within'' -- has been mysteriously 
concealed in plain sight, yet it spins in virtually every slender-body organism: 
in falling cats, crocodilians spinning to subdue their prey, 
rolling fruit-fly larvae, circumnutating plants
and even in some of our own body movements.
Flying somehow under the radar of our cognition, in recent years the wheel-within also tacitly entered the field of soft robotics, finally opening our eyes for its ubiquitous role in Nature. 
We here identify its underlying physical ingredients, 
namely the existence of a neutrally-stable, shape-invariant and actively driven elastic mode.
We then reflect on various man-made realizations of the wheel-within
and outline where it could be spinning from here.
}

The wheel and axle as pinnacles of early human technology are still at the core 
of many of our mechanical devices.
In an abstract sense, the wheel-axle system is an object exhibiting a single, cyclic 
degree of freedom that in spite of its internal rearrangement keeps its outer shape constant, 
i.e.~it moves within its "own skin". This "iso-skinning" feature is exceptionally useful 
as the wheel-axle can be placed within fixed shape enclosures and combined predictably 
with other elements.
Since the earliest days of biology, people have 
wondered whether the wheel-axle principle has been discovered
during evolution and if not, why so. 
The usual suspects for the wheel's (apparent) absence range from its impracticability 
in absence of high quality roads to its difficult discoverablility by small mutations
-- Nature being unable to overcome  "Mount improbable"\cite{Dawkins}.
These are some of the echoing thoughts of self-confident (technological) teenagers -- us, the humanity -- 
reflecting about the seeming shortcomings of their parent.  
Could there be something we are missing here?   It might come as a surprise that Nature found its own version -- and in fact
a technologically superior generalization -- 
of a "cyclic isoskinning" device: the wheel-within. 

Recent work\cite{Drosophila_Nature,Drosophila_prep} 
describes a peculiar motion 
of fruit fly larvae on surfaces. The larva bends its body into a half-doughnut shape
and by contracting its circumferential body muscles in a cyclic manner generates a rotation 
around the curved body axis to finally engage in a rolling-like mono-wheel propulsion. 
The seemingly quirky curiosity turns out to be only one instance of something 
much more universal,
which we coin Nature's wheel-within.
A much earlier encounter dates back to James C. Maxwell, 
who repeatedly 
threw his cat from a (ground floor) window 
and began wondering how it always lands safely "with its feet down"\cite{catbook}. 
It took a century of scientific debate 
to realize that -- similar to the larva -- during the fall, the cat body forms a half-doughnut 
and via differential muscle contraction manages to reorient its 
feet downwards\cite{Marey_FallingCat,Kane_FallingCat,Essen_Cat,Galli}. 
. 

\begin{figure*}[t!]%[b!]
	\centering
	%\vspace{4cm}
	\includegraphics[width=\linewidth]{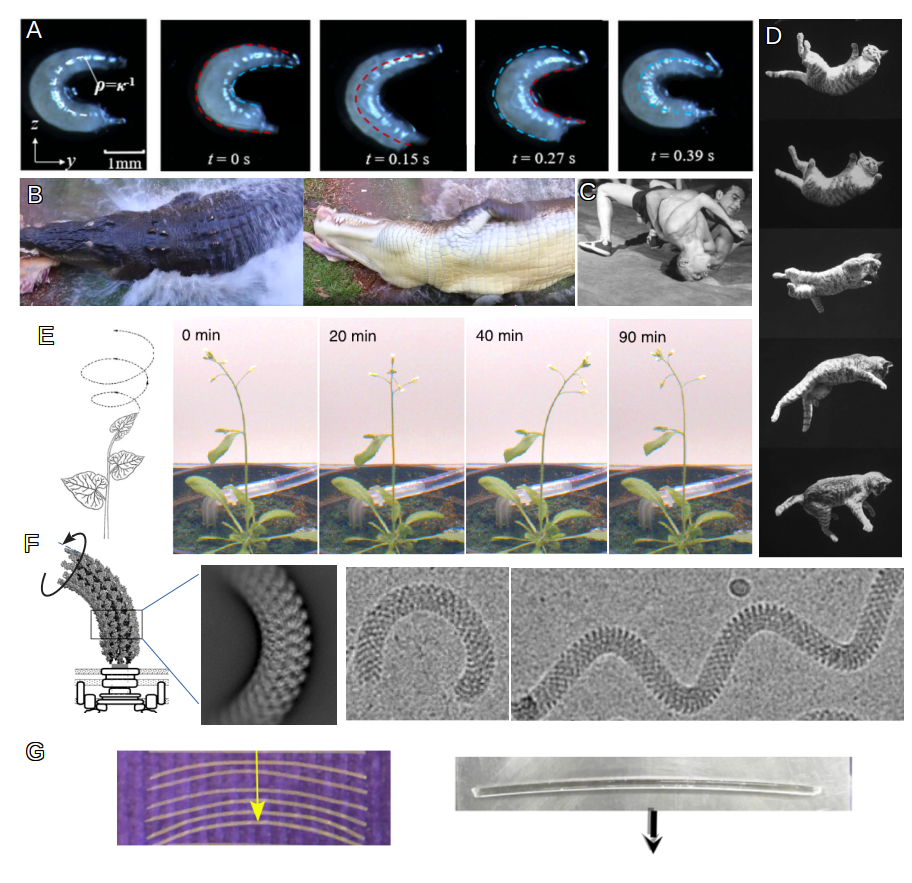}%_new_1404}
	\caption{\label{fig1} {\bf The "wheel-within" is a universal tool of Nature and utilized by 
	many slender organisms in fauna and flora:}
		(A) A Drosophila larva rolling on a substrate\cite{Drosophila_prep}.
		(B) A crocodile's death roll when catching a prey\cite{croco_youtube}.
		(C) A wrestler performing the ``bridge'', also called ``upa'' in Brazilian jiu-jitsu\cite{Bridge}.
		(D) A falling cat, spinning around to land on the feet\cite{catbook}.
		(E) Circumnutation, i.e.~spinning around the gravity axis during growth,
		of {\it Arabidopsis thaliana}\cite{Agostinelli_nutation,Mugnai_nutation}.
		(F) The flagellar hook: the left panel shows a sketch of the hook on top of the rotary motor;
		the other panels show cryo-EM pictures (modified from Ref.\cite{Bact-Hook}). 
		(G) Two artificial, self-organized wheels-within (``fiberboids''): 
		on the left a stroboscopic image of a spaghetti driven by osmosis to
		roll on a humidified kitchen towel, on the right a 
		thermally driven PDMS fiber rolling on a hot substrate\cite{Bazir,Baumann}. 
		}
\end{figure*}

In the following, after familiarizing us with more examples from Nature, cf.~Fig.~\ref{fig1},
we will dissect what they all have in common. We identify the three main physical ingredients
of this unexpected motion: first, the existence of a neutraly-stable, or zero-elastic energy mode; 
second, the shape-invariance or ``iso-skin'' property of this mode in analogy to a wheel;
and finally a suitable, dissipative or active, coordinated or self-organized driving
process of this mode, with 
the principle of ``incompatible equilibria'' or ``dynamic frustration''
being an especially efficient realization.
After explaining the concept, we review the many -- sometimes only partial -- realizations 
of the wheel-within in soft robotics for variously shaped actuators and motors. 
We classify the different occurrences of the wheel-within 
concerning their degrees of freedom and number
of ``wheel-modes''.
Finally we outline possible future directions and unravel the hidden possibilities of the wheel-within.

\section*{The wheel-within in Nature}

\subsection{}\hspace{-2.5mm}
Reflecting on the motion of fruitfly larvae and cats
as shown in Fig.~\ref{fig1},
one quickly realizes a general motif behind: 
the one of a slender object, whose cross-sections contract circumferentially, 
keeping the overall shape (roughly) invariant. 
This motif is found across scales and all over the plant and animal realm: 
similar motion can be observed on the macroscale, for instance, crocodiles spinning 
in the water after having caught their prey
-- the so-called ``crocodile's death roll''\cite{Fish_croco} -- 
or in human pole jumpers or wrestlers. 
Examples from the micro- and nanometric scale
include  microswimmers such as  spirochetes\cite{SpirchetteREVIEW,Wolgemuth}  
(the causative agent of lyme's disease), 
as well as the 
-- rotary-motor-driven -- shape-invariant spinning of the bacterial flagellar 
hook\cite{Namba-1,Bact-Hook,Namba-2}.
In the plant realm the wheel-within hides in a process called ``circumnutation''
-- a phenomenon already described by Charles Darwin\cite{Darwin}:
the spinning of a growing plant around the gravity axis
due to radially asymmetric osmotic swelling and growth rates\cite{Agostinelli_nutation,Mugnai_nutation}.
Some of these examples are illustrated in Fig.~\ref{fig1}. 
The wheel-within enables Earth's flora and fauna to perform various tasks:
to roll on rigid substrates (larvae), 
to swim through viscous fluids (spirochetes) and even to inertially reorient in ``empty space''
(falling cat) and as a clever means to transmit torque around a corner (flagellar hook),
see table \ref{tab_Nature}.
Different -- and somewhat unrelated -- explanations have been put forward
for each of these phenomena. In the physics literature, 
the cat-spinning was treated in the abstract framework of geometric phases 
and an-holonomy\cite{Wilczek_Gauge}, with the notable exception of a 
short 
note by Lecornu \cite{Lecornu_FallingCat} already identifying the cat with a spinning torus. 
Plant circumnutation was rationalized with mechanical models coupled to 
growth\cite{Agostinelli_nutation}. In view of the striking commonalities, however,
it becomes pertinent to identify the unifying concepts that could serve us as a framework 
transferable to man-made machines.

\begin{figure}[b!]
    \centering
    %\vspace{4cm}
    \includegraphics[width=\linewidth]{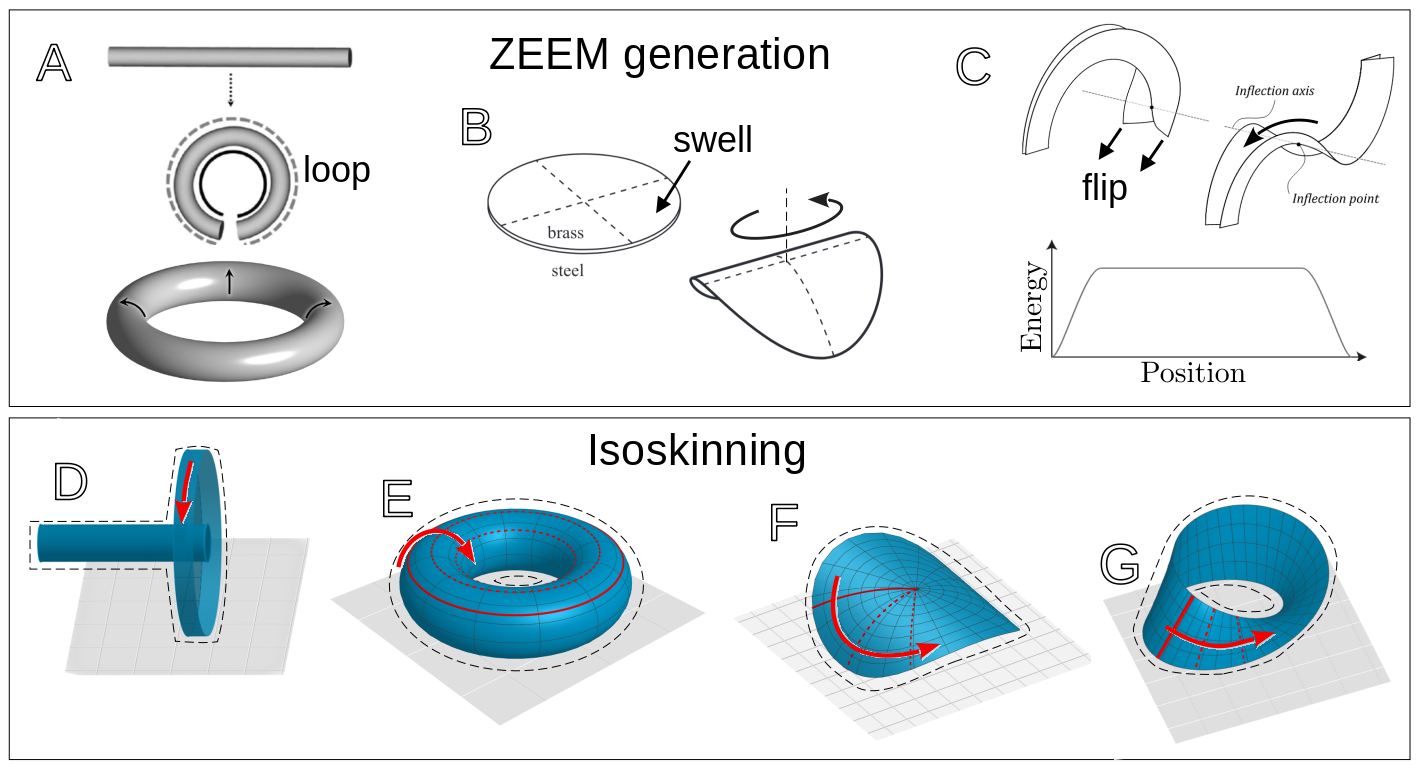}
    \caption{\label{fig2} A-C {\bf Neutrally stable/Zero-elastic energy modes (ZEEM)}.
    D-G {\bf Isoskinning objects with neutral modes}.
    (A) A torus formed by bending a rod and glueing the ends together 
    as the simplest structure displaying a circular ZEEM\cite{Baumann}.
    (B) Shell structure made of two different layers\cite{guest_zero-stiffness_2011}. 
    (C) Neutrally stable
    mode in a ``kinked'' arc-shaped shell\cite{Kok_Herder, VdLans_Radaelli}: 
    the inflection point can be moved along the crest 
    without energy cost, see the flat energy region in the sketch below.
    (D) The rigid wheel-axle system: the wheel turns (red arrow) round the axle  
    while keeping the outer skin (dashed) invariant.
    (E) A neutrally stable torus rotating around its curved centerline.
    (F) A neutrally stable self-buckled shell reorienting by keeping its outer skin fixed.       
    (G) A neutrally stable twist deformation of a M\"obius strip 
    moving along its center-line within the same skin. 
    }
\end{figure}

Initially unaware of Nature's wheel within, some time ago we created a 
family of peculiar elasto-dynamic engines -- the toroidal ``fiberdrive''\cite{Baumann} 
and its open (i.e.~not closed to a torus) fiber analogue, the ``fiber-boids''\cite{Bazir}. 
Fiberboids, see Figure \ref{fig1}G, are macroscopic polymeric fibers 
with circular cross-section that when placed on a surface subject 
to an energy-matter flux (e.g.~a thermal or humidity gradient)
start to roll along the surface.
The fiberdrive, in turn, is a fiberboid closed to a torus, see Fig.~\ref{fig2}A and \ref{fig3}B, 
that if driven rotates in the poloidal direction (i.e.~every cross-section turns around the centerline). 
Playing with these strangely counterintuitive %, shape-invariant kinematics of these 
minimalistic engines, and at the same time seeing the rolling larvae, 
strangely resonates and after some thought strongly suggests that one is seeing instances of the same shape-invariant kinematics and dynamics.   

\section*{Active, neutrally elastic, isoskinning modes}

\subsection{}\hspace{-2.5mm}
So, what are the main physical ingredients for the wheel-within to run?
First, there is the existence of a
{\bf neutrally stable mode}, or {\bf  zero-elastic energy mode (ZEEM)}.
In many physical systems, it is a common phenomenon 
that bi- or multi-stability occurs,
meaning the system displays several energy minima.
Note that in elastic systems, such minima typically imply different shapes. 
If in addition these minima lie continuously along a curve at equal energy one calls the system elastically ``neutrally stable''. 
As a consequence, it costs no energy to move along this path, 
implying in turn a continuous deformation mode at no energy cost.  
Because there is (in the idealized case) no resistance to an applied force, 
in mechanical engineering such elastic systems are  also
called ``zero-stiffness structures''\cite{schenk_zero_2014}, 
exhibiting a 
 ZEEM\cite{Baumann}.

\begin{table*}
\centering
\begin{tabular}{|c|c|c|c|c|c|}
\hline 
{\bf System (Nature)} & {\bf ZEEM} & {\bf Number DoFs} & {\bf Origin of Drive} & {\bf Type of Drive} & {\bf Effect/Use}\tabularnewline
\hline 
\hline 
Cat\cite{catbook}/Crocodile\cite{Fish_croco}/Wrestler & emergent  & 2 & neuromuscular & internal & reorientation in 3D\tabularnewline
\hline 
Rolling Larvae\cite{Drosophila_Nature,Drosophila_prep} & emergent & 2 & neuromuscular & internal & rolling + translation\tabularnewline
\hline 
Spirochete\cite{SpirchetteREVIEW,Wolgemuth}  & emergent & 2 & molecular motors & internal & swimming\tabularnewline
\hline 
Circumnutating Plant\cite{Darwin} & intrinsic  & 2 & dynamic osmotic stress & internal & translation + rotation\tabularnewline
%\hline 
% &  &  &  &  & \multirow{1}{*}{}\tabularnewline
%\hline 
% &  &  &  &  & \tabularnewline
\hline 
Bacterial Flagellar Hook\cite{Namba-1,Bact-Hook,Namba-2} & intrinsic  & 1 & rotary motor & external & torque transmission\tabularnewline
\hline 
\end{tabular}
\caption{\label{tab_Nature}
{\bf Examples from Nature.} The ZEEM can be emergent 
(i.e.~the curvature is created concomitantly)  or intrinsic. The number
of degrees of freedom (DoFs) in the emergent cases as well as in plants is two, since not only the
orientation of curvature but also its absolute value are variable. The flagellar hook
has only one degree of freedom. The other rows discuss the driving mechanism, 
whether the motion is self-organized ore externally driven, and what the motion is typically used for.
All examples
have (albeit some only approximately) the isoskinning property. 
}
\end{table*}

Some example structures are shown in Fig.~\ref{fig2}A-C.
Fig.~\ref{fig2}A shows the fiberdrive, 
a torus displaying a ``circular ZEEM''\cite{Baumann}:
rotating a cross-section in the poloidal direction does not cost energy.
The underlying reason is that, due to the topological constraint of the torus shape,
the absolute value of the (centerline) curvature is fixed, but its orientation
with respect to a cross-section is free and constitutes a zero-elastic energy mode. 
Fiberboids\cite{Baumann,Bazir} -- open fibers -- can be considered as parts of a torus
with their curvature being induced by the same energy-matter flux that drives them. 
Other examples include bilayered shells, cf.~Fig.~\ref{fig2}B,
as well as edge-crumpled sheets and several variants of M\"obius strips.
A very illustrative example for a  ZEEM is shown in Fig.~\ref{fig2}C 
and was introduced in Refs.\cite{Kok_Herder, VdLans_Radaelli}.  
The structure is an arc-shaped shell, where an  inflection point can be moved freely
along a one-dimensional neutrally stable path, since the energy is constant (see sketch)
except at the ``ends'', where it decreases
as the structure prefers a shape without inflection point there. 
While the other examples display circular ZEEMs, as required for a wheel,
the last example is a ``translational ZEEM''.

\begin{figure}[b!]
    \centering
    %\vspace{6cm}
    \includegraphics[width=\linewidth]{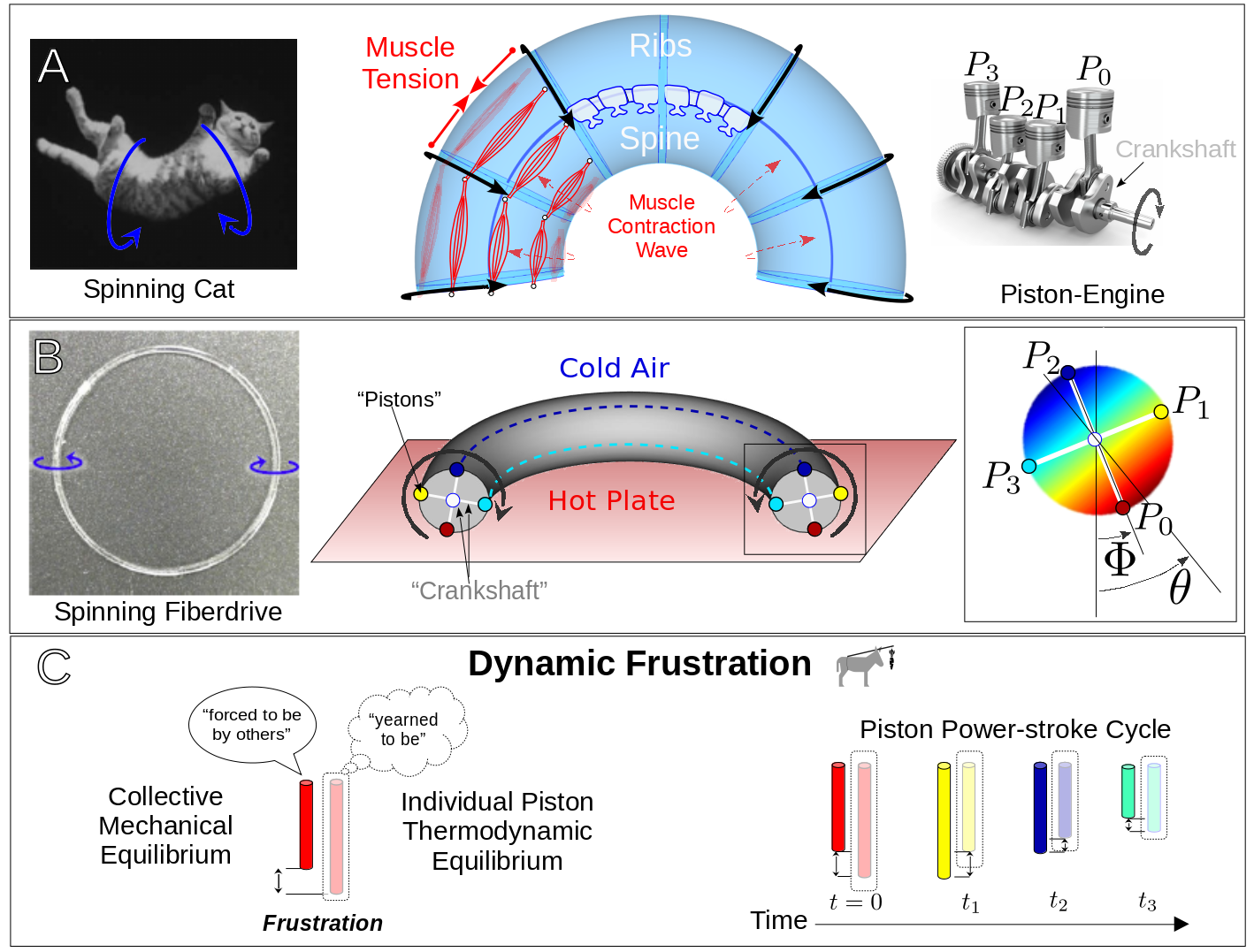}
    \caption{\label{fig3} {\bf Active driving, crankshaft analogy and dynamic frustration.}
    (A) A falling cat, spinning to get on its feet, seen as part of a torus. 
    The analogy to the piston engine: the spine/ribs correspond to the crankshaft and 
    the contractile muscles (which are actuated by the animal) to the pistons.
    (B) In the fiberdrive, a closed torus heated from below, one can consider parts of material,
    e.g.~as indicated, as the pistons. The crankshaft arises due to the material's connectivity.
    (C) Dynamic frustration/incompatibility of equilibria: every piston has a current length (left)
    that differs from its thermodynamically preferred length (right) at the given temperature.
    This  leads to a situation where  every piston is ``frustrated'', creating an overall torque. 
    }
\end{figure}

Inspecting the examples from Nature again,
there the ZEEM is intrinsic for circumnutating plants (gravity buckling)
and the bacterial flagellar hook (incompatible strains),
but it can also be emergent, i.e.~self-organized under intrinsic forces,
as in the spinning cat and spirochetes, cf.~table \ref{tab_Nature}.
All examples from Nature, except the flagellar hook, have two degrees of freedom:
the one associated to the ZEEM (the rotation around the curvature centerline)
and the absolute value of the curvature (which will be identified later as a ``modulator mode'', see
Fig.~\ref{fig5}). 
In that sense the examples from Nature correspond most closely 
to the fiberboid (Figure \ref{fig1}G), 
with most other man-made examples so far (torus, M\"obius strip, etc.) 
having only one degree of freedom  (see later and Fig.~\ref{fig5}).

% THE BOX -------------------------------------------------------------------------------------------------------
%
%
%\begin{minipage}[t]{\textwidth}
\begin{figure*}[t!]
\begin{mdframed}[backgroundcolor=myframecolor,linecolor=myframecolor,fontcolor=black,
	%innertopmargin=9.3pt,innerrightmargin=11.4pt,innerbottommargin=12pt,innerleftmargin=11.4pt,
	linewidth=0pt,userdefinedwidth=\textwidth]% 155pt
	\textcolor{black} 
{\bf Box 1: Driving the wheel-within: Piston-crankshaft analogy.}
\vspace*{1mm}

\begin{minipage}{.485\textwidth}
%\centering
As explained in Fig.~\ref{fig3}, the fiberdrive has a formal analogy 
with a piston engine. We consider here a toy model 
of 4 sub-fibers ("pistons") arranged at 90$^\circ$ orientations 
running at constant distances across each cross-section around the torus.
They will be referred to as pistons, since their length
changes coupled with the fiber geometry lead to a collective rotation
of the whole device, the fiber geometry itself acting as an equivalent
of the crankshaft 
by enslaving the pistons to share a common 
orientation angle $\Phi(t)$ of the cross-section with respect to
the outer frame. 
Each piston fiber ($k=0,..,3$) is characterized via\\
a) the time-dependent temperature 
$T_{k}(t)$,
implying via an equation of state their \textbf{mechanically preferred}
piston extensions $x_{k}(t)$. \\ %, at any time $t$.\\
b) the \textbf{thermodynamically preferred} piston temperatures 
\begin{equation}
T_{k}^{ext}\left(\Phi\right)  =\overline{T}\left(1+a_{T}\cos\left(\Phi+k\frac{\pi}{2}\right)\right),
%\text{ (Bath geom.)}
\end{equation}
as set by the external baths and %(at given angular orientation $\Phi+k\frac{\pi}{2}$),
characterized by  the mean temperature $\overline{T}$ and $\Delta T^z_{ext}=2\overline{T}a_{T}$
its variation in $z$-direction , perpendicular to the torus plane. \\
c) the \textit{mechanically imposed} piston extensions 
\begin{equation}
X_{k}\left(\Phi\right)  =L\left(1+a_{X}\sin\left(\Phi+k\frac{\pi}{2}\right)\right),
\end{equation}
as enforced by the action of the \textbf{crankshaft} geometry, where $L=2\pi / \kappa$ is the 
centerline length of the closed torus with (large-circle) curvature $\kappa$. 
This  curvature  and the imposed piston extension are related 
via the strain $\epsilon=\frac{L a_X}{L}\simeq\kappa a$, where $a$ is the (poloidal) radius of the torus - also assumed to coincide with the lateral size of the pistons.

Note that both (1) and (2)
are explicit functions of a single \textbf{collective variable} - the angle $\Phi$. 
They both are constrained by the geometry of the device  and
their mismatch (here a 90$^\circ$ phase shift cf.~$\cos$ vs.~$\sin$) 
together with the inability of the system to match both 
is exactly what drives the engine.

There are two types of dynamic relaxation processes, a thermodynamic and a mechanical 
one. The \textbf{thermodynamic relaxation dynamics}
is given by a Newton's law of cooling
% (``https://en.wikipedia.org/wiki/Newton's\_law\_of\_cooling'')
\begin{align}
\dot{T}_{k}\left(\Phi,t\right)  =\gamma\left(T_{k}^{ext}\left(\Phi\right)-T_{k}\left(\Phi,t\right)\right), %\text{ (Newton law of cool.)}
\end{align}
where we assume that the pistons can be seen as having a uniform 
temperature\footnote{This is the case for small Biot number $Bi=ha/k_{p}\ll1$, 
where $h$ is the heat transfer coefficient, $k_{p}$ the piston's
thermal conductivity and $a$ its diameter. 
The relaxation rate $\gamma$ can then be expressed in terms 
of the specific heat capacity $c_{p}$, the mass of the piston $m_{p}$ and the 
effective area $aL$ for heat exchange, as $\gamma=\frac{haL}{m_{p}c_{p}}$.}.
$\gamma$ is a relaxation rate and its inverse $\tau=\frac{1}{\gamma}$ 
represents the typical thermal relaxation time for the pistons to equilibrate 
with their environment. 

The temperature of the pistons and their extensions result in a force
that originates from the \textbf{equation of state} specific for
the piston material. We assume the simplest expansion of the free energy
for small temperature and length changes
\[
\mathcal{F}\left(T,X\right)\approx\mathcal{F}_{0}\left(T\right)+\frac{YLa^{2}}{2}\left(\alpha\left(\overline{T}-T\right)-\frac{L-X}{L}\right)^{2}\,.
\]

\end{minipage}%-----------------------
\begin{minipage}{0.03\textwidth}
\hspace*{.1cm}    
\end{minipage}%-----------------------
\begin{minipage}{0.485\textwidth}

Apart from an extension independent part, %this free energy
it has a linearly elastic part that couples the thermal prestrain with the
actually realized strain. 
The  thermal prestrain depends on the thermal expansion
coefficient $\alpha$ and the temperature difference to the reference
temperature, $\overline{T}$, corresponding to vanishing prestrain.
The Young's modulus $Y$ is assumed to be temperature independent.
Differentiating the free energy with respect to the piston extension 
yields
the piston force
\begin{equation}
f_{k}  \left(\Phi,t\right)=Ya^{2}\left(\frac{L-X_{k}\left(\Phi\right)}{L}-\alpha\left(\overline{T}-T_{k}\left(\Phi,t\right)\right)\right)\,. %\,\text{(Thermodyn eqn of s.)}
\end{equation}
Noting that the crankshaft
constrains the system to only a single degree of freedom, the collective angle $\Phi$, we can write down a Lagrangian
\[
\mathcal{L}\left(\Phi,\dot{\Phi}\right)=\sum_{k=0}^{3}f_{k}X_{k}\left(\Phi\right)+\Phi M_{ext}
+\frac{I}{2}\dot{\Phi}^{2}
\]
where the first term describes the potential energy (of piston extensions),
the second term the effect of a possible external torque and 
the last term rotational kinetic energy (with $I$ the
effective moment of inertia), which will be neglected in the following for slow turning.
Forces originating from mechanical dissipation, i.e.~friction, can be accounted for by
introducing the Rayleigh dissipation functional $\mathcal{R}=\frac{1}{2}\xi\dot{\Phi}^{2}$.
The mechanical force balance, given by the \textbf{Rayleigh-Lagrange dynamics}, 
\[
\frac{\partial\mathcal{L}}{\partial\Phi}-\frac{d}{dt}\frac{\partial\mathcal{L}}{\partial\dot{\Phi}}=\frac{\partial\mathcal{R}}{\partial\dot{\Phi}},
\]
then  simply yields
\begin{equation}
\sum_{k=0}^{3}f_{k}\left(\Phi,t\right)\frac{\partial X_{k}\left(\Phi\right)}{\partial\Phi}+M_{ext}  =\xi\dot{\Phi}\,. %\,\text{(Euler-Lagrange-Rayl. eqn.)}
\end{equation}

Using as an ansatz for the actual piston temperature
\begin{equation}
T_k(t) =T_0+T\cos\left(\Phi+\theta+k\frac{\pi}{2}\right)\,,
\end{equation}
with a to be-determined phase angle $\theta$,
equations (1)-(5) are easy to solve  in the steady state
by replacing $\Phi\rightarrow\omega t$ with a finite angular velocity $\omega$. 
The resulting \textbf{motor-relation} (torque-angular velocity relation) reads
\begin{equation}
\left(\frac{\omega^2}{\gamma^2}+1\right)\left(\xi\omega-M_{ext}\right)=M_{drive}\,.
\end{equation}
The driving torque is given by
\begin{equation}
M_{drive}=\left(L a^3\right) Y \kappa \left(\alpha \Delta T^z_{ext}\right)
\end{equation}
%=2 \alpha a_{T}\,a_X  \overline{T} Ya^2 L
and has a simple interpretation:
up to a geometrical factor it is given by the stiffness, 
the geometry-imposed curvature and the 
thermal expansion-induced driving strain. 
For small drive (and no external torque), 
the angular velocity is simply given by
$\omega=\frac{M_{drive}}{\xi}$ and the  phase angle
amounts to $\theta=- \frac{\omega}{\gamma}$.
\end{minipage}
\end{mdframed}
\end{figure*}
%\end{minipage}
%
%
%-----------------------------------------------------------------------------------------------------------

The second major ingredient for the wheel-within to work is the
{\bf  Isoskinning} (or Iso-surface) property: 
to be a true wheel-within,  the motion of the body must keep its enclosure's surface (its ``skin'') 
invariant, see  Fig.~\ref{fig2}D-G. %, to which we refer to as ``isoskinning''.
Of course in Nature -- thinking of the cat -- this may just be approximately realized.
The simplest example is obviously the classical, man-made, rigid wheel, 
Fig.~\ref{fig2}D.
The most elegant wheel-within is probably the torus-shaped fiberdrive\cite{Baumann}, 
see Fig.~\ref{fig2}E, where every cross-section rotating at constant speed 
leaves the whole object shape-invariant.
More examples include bent sheets, see Fig.~\ref{fig2}F  
and realized in Ref.\cite{hamouche_multi-parameter_2017}, 
as well as M\"obius strips, see Fig.~\ref{fig2}G and realized in
Refs.\cite{Moebius_old,Moebius3,Moebius_new} (see also Fig.~\ref{fig4} for more examples 
of man-made realizations). 
In all the sketches of Fig.~\ref{fig2}D-G, the dashed curve describes 
the invariant ``skin'' of the object and the red arrow indicates the ZEEM direction. 
It should be noted that, except for the trivial case of a purely rigid rotation, 
isoskinning is \textit{never} an isometry of the body as the distances between 
points are periodically changing in the body frame.
Interestingly, in case of the falling cat a counter-rotation of the skin itself exists 
to conserve the total angular momentum -- in fact this ``skin rotation'' is the very purpose 
for the cat using the wheel-within and to land on its
feet\cite{Marey_FallingCat,Kane_FallingCat,Essen_Cat,Galli}.

The third and last ingredient of the wheel-within is an {\bf Active drive}:
to make the wheel turn, the neutral mode having the isoskinning property 
has to be driven by a nonequilibrium process coupling to elastic stresses and strains. 
Many different driving mechanisms appear possible and some of Nature's realizations include
active (ATP-consuming) muscle contraction in the animal realm, cf.~Fig.~\ref{fig3}A,
or osmotic stresses in plants.
For man-made systems, a temperature difference/heat flux
can be used as in the fiberdrive\cite{Baumann}, 
a polymeric macroscopic torus heated from below, 
as shown in Fig.~\ref{fig3}B and with the mechanism detailed in box 1. 
Other examples include hygroscopic (de)swelling\cite{Ghatak1,Bazir}, 
i.e.~a solvent enters/leaves a network-based material, 
or light. The latter can drive the ZEEM either indirectly via 
local heating\cite{AhnElastomer,lightdrivenTorus},
or directly, e.g.~via cis-trans-isomerization of light-sensitive molecules like azobenzene
that are embedded in a matrix\cite{Choi_Kim_azobenz_helix_fibers}.

In spite of this plethora of possible driving mechanisms, there is a serious catch 
that whoever practically tries to induce the isoskinning motion 
along a ZEEM quickly can confirm: not all types of non-equilibrium stress generation 
will induce the desired motion. 
In fact, most will do nothing at best, merely deforming and more often than not irreversibly 
destroying the sample. However, there is a systematic solution for efficient driving:
the principle of ``incompatible equilibria'' resulting in a steady state 
of ``dynamic frustration''.

\section*{Incompatible equilibria and Dynamic frustration}

\subsection{}\hspace{-2.5mm}
Rotation of the wheel-within needs some 
form of highly coordinated, geometrically orchestrated generation 
of stresses and strains that couple out of phase to generate torques. 
That is where a phenomenon that we might call dynamic frustration comes into the game.

\begin{figure*}[t!]
    \centering
    %\vspace{4cm}
    \includegraphics[width=\linewidth]{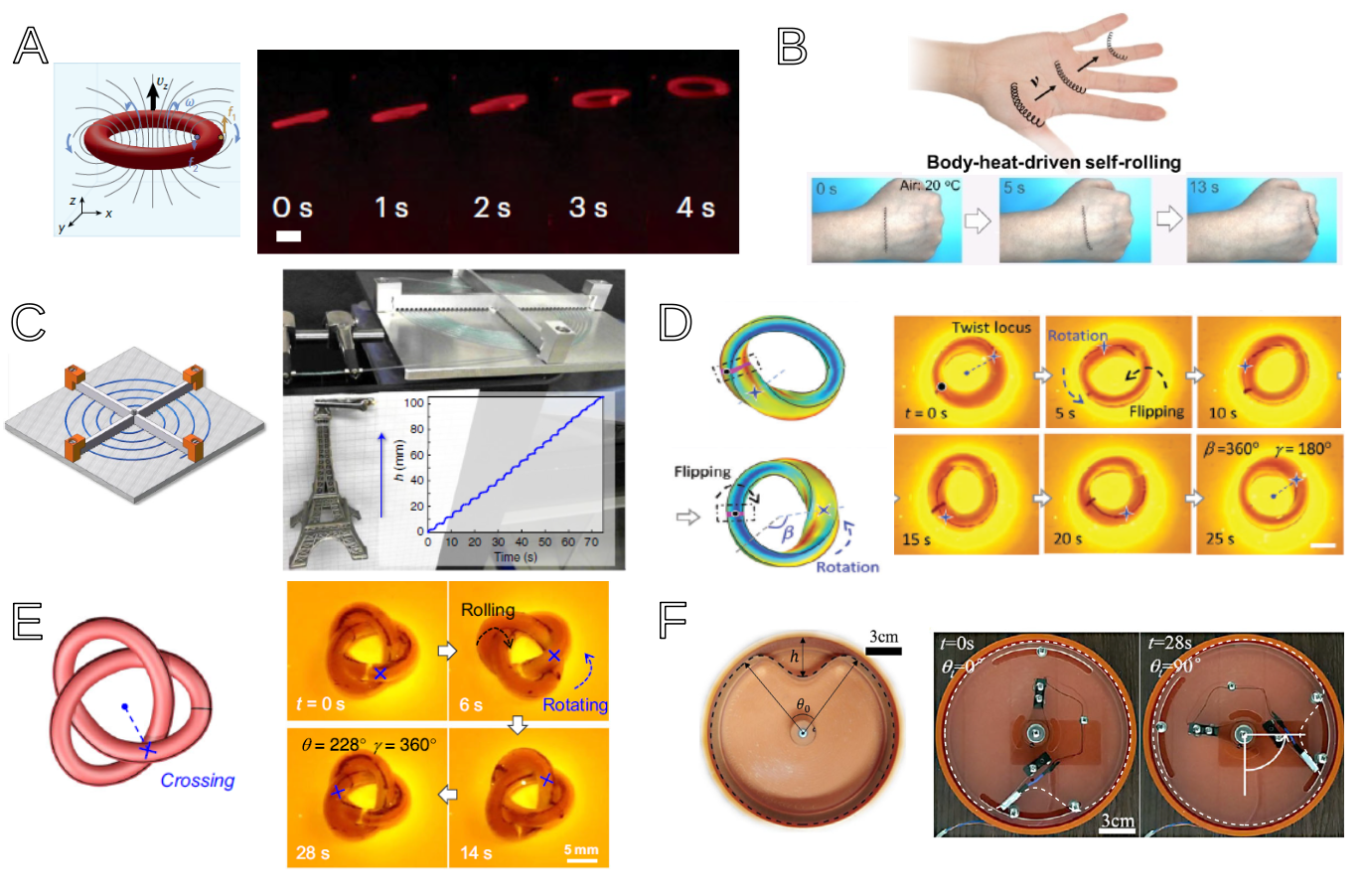}
    \caption{\label{fig4}{\bf Man-made examples  of the wheel-within.}
    (A) A photo-thermally driven fiberdrive (torus) swimming in a liquid
    in the low Reynolds 
    number regime\cite{lightdrivenTorus}.
    (B) A (helical) fiberboid utilizing body heat to roll along a human arm\cite{bodyheat}.
    The helical shape slightly breaks the perfect ZEEM, but also has advantages 
    like a better adaptation to the surrounding. 
    (C) An Archimedean spiral can be seen as several fiberdrives mounted in series, 
    adding up the torques they can exert to lift a weight\cite{Baumann}.
    (D) A  M\"obius strip made of a composite material (anisotropic hydrogel 
    with plasmonic nanoparticles) that rotates under static light illumination\cite{Moebius3}.
    (E) A ``knotbot'', performing two modes of motion:
    poloidal (``rolling'') and toroidal (``rotating'') under static light illumination\cite{knotbots}.
    (F) A confined blister that rotates when heated along its circumference; 
    modified from Ref.\cite{blister}.
    }
\end{figure*}

To illustrate the concept intuitively, we may remark that many of the 
biological and man-made examples can be
understood by thinking of a rotary piston-engine, see Fig.~\ref{fig3} and box 1,
with their pistons performing thermodynamic power-stroke cycles. 
Much like in a car engine, the pistons are not independent but rather rigidly coupled 
via some form of constraint that we might call the "crankshaft" in analogy to classical engines. 
In a falling cat, cf.~Fig.~\ref{fig3}A, the crankshaft is constituted of the cats incompressible spine, 
together with its radial rib-skeleton connecting up to the muscles which act as the pistons.
In the toroidal fiberdrive, cf.~Fig.~\ref{fig3}B, the crankshaft is emulated by approximate 
mechanical constraints of the cross-section's circularity and the centerline incompressibility. 
This virtual crankshaft enslaves the contracting pistons 
-- the longitudinal muscles or the actuating material sections --
to share one or two common degrees of freedom, forcing them to synchronize their action.   
An idealized cat has two degrees of freedom, the amplitude and the phase angle of its body curvature,
while a toroidal fiberdrive has only one degree of freedom (the phase), since the magnitude 
of the curvature is rigidly maintained by the closure constraint of the torus. 
The fiberboids, linear rolling fibers, can be viewed as parts of a torus
and are the  analogue of the cat in the sense that they have again two degrees of freedom,
phase and magnitude of curvature, the latter
emerging from the dissipative coupling of the fiber with the planar substrate.
In box 1 and Fig.~\ref{fig3}C
we explain a simple one-dimensional version of the fiberdrive, based on 4 coupled pistons,
in more detail, including the specific driving mechanism, the equation of state 
that couples the drive to the pistons,
and the resulting (steady-state) dynamics. 

Importantly, beyond the coupling via the ``crankshaft'', rotary motion needs 
a phase shift of the thermodynamic and mechanical equilibrium positions,
which induces what we call ``dynamic frustration''. 
For the fiberdrive, a torus lying on a hot plate, 
the phase shift is 90$^\circ$: the geometry induces in-plane strains (due to the torus' curvature)
while the thermal gradient induces (thermal expansion) strains in the direction normal to the plane.
Other phase shift values are of course possible but typically less efficient.
Importantly, this phase shift renders the two equilibria -- thermodynamic vs.~mechanic -- mutually incompatible.
This incompatibility can hence be purposefully introduced to prevent 
the existence of any static equilibria, enforcing a constant state of motion of the system, 
much like the allegorical donkey that perpetually chases a carrot 
attached via a rigid pole to its back.
 
Finally, a more subtle point is the relation between the thermodynamic and mechanical variables,
that closes the system in the physical sense. 
In a purely physical system this is simply the piston's equation of state (cf.~box 1),
that relates temperature -- or chemical composition, etc., depending on the specific drive -- 
to its length and generated force. 
In biological systems, like the spinning cat or rolling larvae, this closure relation is more complex
and sometimes difficult to specify.
For the cat it results from neuromuscular feedback closely related to the organism's ability to perceive 
its own body shape and space orientation and to dynamically steer muscle tension 
in accordance to them, in order to reach the goal of the created motion, landing on its feet.

\section*{Man-made wheel-within}

\subsection{}\hspace{-2.5mm}
Without noticing its natural existence and fully grasping its unifying concept, 
the wheel-within recently entered our technology.
Examples are shown in Fig.~\ref{fig4}, 
and table \ref{tab_manmade} provides a rough classification.
As our focus lies on the physical principles,
we do not give an exhaustive review here and
refer to Ref.\cite{ZEEM_review} for an overview of realizations of the wheel-within
from the materials science point of view --
focusing on liquid crystalline elastomers (LCEs) as an especially versatile material --
and to Ref.\cite{robotics_review} for a review  
from the viewpoint of autonomous robotics.  

Starting with the torus geometry, it was introduced 
using basic polymeric materials (nylon, PDMS), on the macroscopic
scale (cm sized) and with thermal drive (heating plate) in Ref.\cite{Baumann}. 
The geometry can be modified, e.g.~to twisted rings\cite{twisted_ring}
and to Archimedean spirals\cite{Baumann}, see Fig.~\ref{fig4}C.
The latter can be regarded 
as several concentric tori put in series,
allowing to add up and hence magnify the created torque. 
More importantly, the driving mechanism was replaced by 
light (photothermal)\cite{lightdrivenTorus,pneum_torus} and the object could be miniaturized.
Several interesting modes of motion can be achieved with tori,
including translational motion\cite{Baumann}, 
swimming in a viscous fluid, as shown in Fig.~\ref{fig4}A,
and moving along guiding tracks including fibers and interior walls of 
micro-pipettes\cite{lightdrivenTorus}.

There is a plethora of examples for fibers, see also Ref.\cite{ZEEM_review}. 
Interesting early work on a macroscopic PDMS cylinder exists,
where a droplet of solvent was placed on one side, inducing hygroscopic swelling and transient uphill rolling\cite{Ghatak1,Ghatak2}.
Though the motion was not yet perfectly self-sustained, some elements of shape 
invariance and ``dynamic frustration'' emerged there, foreshadowing following developments.
Self-rolling started with Refs.\cite{Baumann,Yu_Cai_fiber,Bazir}
using very basic polymeric materials, while nowadays advanced
materials have been designed, including as examples
light- or thermally powered mono-domain LCE rods\cite{AhnElastomer}
and light-driven core-shell fibers\cite{Yu_Lu_coreshellfiber}.
%explanation seems to be wrong there (deformation of cross-section induces 
%center of gravity shift, which induces ``rolling'') 
Multiple modes of motion have been described on 
different substrates\cite{Zhou_Zheng_multimodal} and
twisted/helical fibers have been created\cite{Jiang_helix}.
The latter can  even roll on a human arm,
utilizing just body heat\cite{bodyheat}, see  Fig.~\ref{fig4}B, 
and helical fibers have been designed to
fulfill basic robotic tasks and navigate in a 2D maze\cite{Zhao_maze}.
%Azobenzene-functionalized LCE-based rolling locomotors with photo-induced helix structures:
Finally, since photothermal activation is often prone to damaging the material,
direct (i.e.~not photothermal) light activation  using photoswitchable azobenzene 
was achieved as well\cite{Choi_Kim_azobenz_helix_fibers}.

The M\"obius strip geometry also allows for ZEEMs, see Fig.\ref{fig4}D,
and several variants, typically photothermally driven, 
have been realized\cite{Moebius_old,Seifert_ribbon,Moebius3,Moebius_new}.
First the light source had to be focused on the locus of the twist regions
and hence had to be continuously moved\cite{Moebius_old} 
-- the cooling time of the material was too long and thermal damage prevented 
``dynamic frustration''. 
Later on, activation by continuous radiation was achieved, first for LCE-based 
Seifert ribbons\cite{Seifert_ribbon}. 
Using PNIPAm-based hydrogels with 
embedded nanosheets (for anisotropic response) and gold nanoparticles 
(for light absorption), finally also M\"obius strips\cite{Moebius3} were realized
that rotate under continuous radiation,
profiting from self-shadowing effects followed by new exposure to light when turning. 
 
 \begin{table*}[h!]
\centering
\begin{tabular}{|c|c|c|c|c|c|}
\hline 
{\bf System (Manmade)} & {\bf ZEEM} & {\bf Number DoFs} & {\bf Origin of Drive} &  {\bf Effect/Use}\tabularnewline
\hline 
\hline 
Torus (fiberdrive)\cite{Baumann,lightdrivenTorus,pneum_torus} %(Fiberdrive)
 & intrinsic & 1 & thermal\cite{Baumann}, photothermal\cite{lightdrivenTorus} &  torque transmission, swimming,\tabularnewline
 &&&& track-based motion\tabularnewline
\hline 
Spiral\cite{Baumann}& intrinsic  & 1 & thermal %\cite{Baumann}  
&  torque generation\tabularnewline
\hline 
Fibers (fiberboids)\cite{Baumann,Bazir,Yu_Cai_fiber,AhnElastomer,Choi_Kim_azobenz_helix_fibers}& emergent & 2 & thermal\cite{Baumann,Bazir,Yu_Cai_fiber}, hygroscopic\cite{Bazir},  &  rolling, translation,\tabularnewline
&   & (1 ZEEM + 1 modulator)  &photothermal\cite{AhnElastomer}, photomechanical\cite{Choi_Kim_azobenz_helix_fibers} 
&  robotic tasks%\cite{Zhao_maze}
\tabularnewline
\hline 
M\"obius/Seifert strips\cite{Moebius3,Seifert_ribbon}  & intrinsic  & 1 & photothermal %\cite{Moebius3,Seifert_ribbon} 
& spinning, track-based motion\tabularnewline
\hline 
Knots\cite{knotbots} & intrinsic  & 2 (2 ZEEMs) & photothermal &  spinning, track-based motion\tabularnewline
\hline 
Sheet\cite{hamouche_multi-parameter_2017}, Blister\cite{blister} & intrinsic  & 1 & thermal  &  spinning, fluid pumping%\cite{blister}
\tabularnewline

\hline 
\end{tabular}
\caption{\label{tab_manmade} {\bf Man-made examples.}
A table analogous to table \ref{tab_Nature} for man-made examples,
discussing whether the ZEEM is emergent or hard-coded in the geometry/topology,  
the number of degrees of freedom (DoFs), the driving mechanism
%whether the motion is self-organized, 
and what the motion can be used for.
In addition, all examples in the table are isoskinning.
}
\end{table*}

A very interesting geometry class with a novel effect is the so-called ``knotbot''\cite{knotbots},
i.e.~a closed knotted filament as shown in Fig.~\ref{fig4}E.  
It displays not only the poloidal ZEEM, similar to the simple (unknotted) torus,
but in addition a second ZEEM: a toroidal motion along the circumference,
to which we will come back below.
 
Finally, actuated sheet geometries have been proposed, 
like the kinked arc-shaped sheet\cite{Kok_Herder}, cf.~Fig.\ref{fig2}C,
that was recently also thermally driven\cite{VdLans_Radaelli}.
Although this example lacks the isoskinning property, it still appears 
very appealing for robotic actuation. 
Neutrally stable modes have been shown to exist in several other slender 2D objects 
like prestressed shells and sheets\cite{pellegrino_iutam-iass_2000,guest_zero-stiffness_2011,seffen_prestressed_2011,schenk_zero_2014}
and some of these modes were actively 
driven\cite{vehar_closed-loop_2004,hamouche_multi-parameter_2017}.
A nice example of a true isoskinning ZEEM was achieved by
confining a sheet in a cylindrical enclosure, yielding a rotating blister\cite{blister}, 
see Fig.~\ref{fig4}F.

\begin{figure*}[t!]
	\centering
	%\vspace{4cm}
	\includegraphics[width=\linewidth]{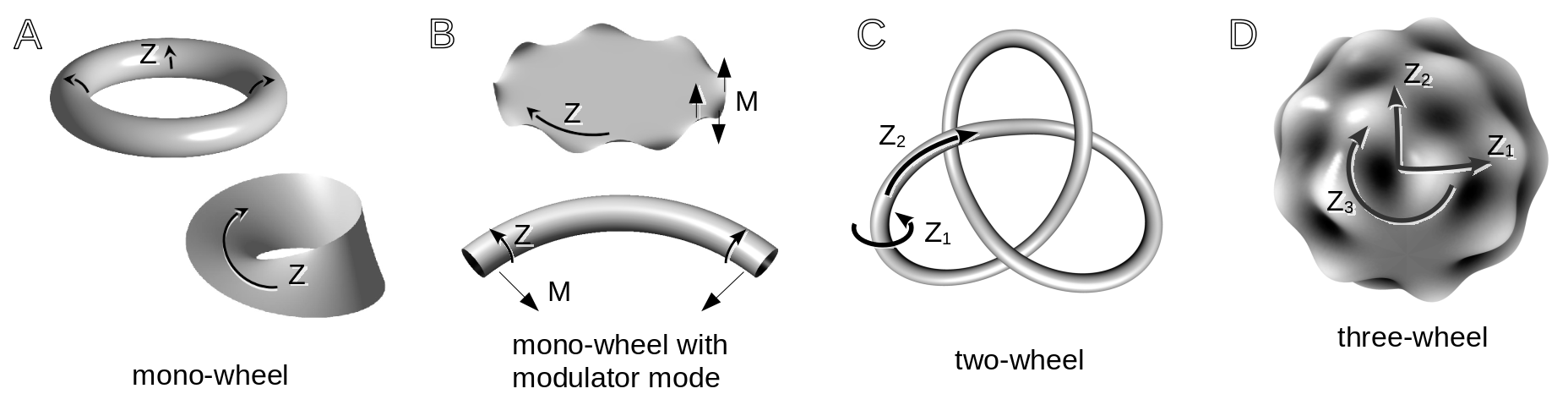}
	\caption{\label{fig5}{\bf Degrees of freedom of the wheel-within family.} 
	A) The fiberdrive (torus) and the M\"obius strip as ``mono-wheels'' with a single ZEEM (Z). 
	B) Fiberboids (open fibers) and crumpled sheets have two degrees of freedom:
	one ZEEM (Z) and one auxiliary bending mode (M) 
	with finite stiffness that can modulate the active dynamics of the ZEEM. 
	All known wheels-within in Nature are of this type, with the exception of the
	flagellar hook. 
	C) Any elastic knot (knotbot) has two ZEEMs: a poloidal ($Z_1$) 
	and a toroidal ZEEM ($Z_2$). 
	D) A raspberry-like deflated and crumpled elastic sphere can in principle
	have three ZEEMs, corresponding to three spatial rotations.}
\end{figure*}

To classify the plethora of geometries proposed so far, it is important to note that
the ZEEMs occurring in the so far created man-made examples 
are all intrinsic: they are built-in either topologically 
(e.g.~in the torus, the M\"obius strips and the knotbots) 
or set by other constraints (spinning spiral, bulged sheets)
or incompatible stresses (for crumpled sheets).
The only exception seems to be the
self-rolling fiber -- and the examples from Nature: here the active, 
neutrally elastic, isoskinning mode is spontaneously emerging 
(bifurcating) from the cylindrically symmetric object in the absence of the drive.

Another interesting classification relates to the number of ZEEMs, see Fig.~\ref{fig5}:
the torus has the poloidal ZEEM that was our
main focus so far, while the M\"obius-like strips have a single toroidal ZEEM, cf.~Fig.~\ref{fig5}A. 
The crumpled sheet and the fiber geometry also have only one ZEEM,
but now this mode can be modulated, see Fig.~\ref{fig5}B: 
the amplitude of the crumpled rim and the curvature of the fiber, respectively, 
depend on the amount of drive.
This ``modulator mode'' makes the fiber geometry so versatile as it can
adapt to the surrounding -- and it hence seems not to be a coincidence 
that the examples from Nature have this modulator mode. 
As seen in Fig.~\ref{fig4}E, the knotbot is again different: it has two ZEEMs,
one in the poloidal and one in the toroidal direction, cf.~Fig.~\ref{fig5}C, 
and hence could be called
a ``two-wheel'' to distinguish it from the ``mono-wheels'' discussed so far.
Finally, a ``three-wheel'' is also possible theoretically, with one potential realization
being a crumpled (e.g.~deflated) spherical shell as shown in  Fig.~\ref{fig5}D.  
In all these examples the ZEEMs correspond to modes 
restoring broken continuous symmetries 
of the underlying manifolds, the symmetries being broken either 
by topological constraints or by bifurcations. 
In this sense, they can be seen as geometric actively driven analogues of 
Goldstone modes\cite{Forsterbook,Lubensky_Goldstone}.
Such modes appear in bulk condensed materials whenever a continuous symmetry 
is broken and correspond to slow (hydrodynamic), low (zero) energy modes 
with classical examples being the rotation of the magnetization or the nematic director 
in a ferromagnetic solid or a nematic liquid crystal, respectively\cite{Forsterbook}. 
Among the Goldstone modes in solids, the ZEEMs  however stand out in that they 
involve a literal matter-flux along the respective directions. Thus, they can be 
seen as solids with one (or respectively two or three, see Fig.~\ref{fig5})
entrapped fluid modes.

\section*{Perspectives: Where the wheel-within is rolling}

\subsection{}\hspace{-2.6mm} 
By creating artificial wheels-within with colleagues in the field we unknowingly 
stepped into Nature's territory, offering a blissful view 
to the mysterious minimalism and efficiency of the natural wheel-within:  
It shows no frictional wear between its components, as it has only a single one. 
It can emerge into and pass away from existence in any slender organism -- 
after all a cat has many more things to do than spinning in the air. 
A snake, for instance, can propagate sinusoidal waves down its body to swim or 
slither on the ground, it can generate propagating kink pairs to 
climb trees\cite{jaynes_Snakes_2020}
and it can spin its wheel-within if it needs so. 
Also a manta-ray can propagate waves along its leaf-like body to thrust forward, 
or it can create an isoskinning mode to reorient its swimming direction.
In this sense any slender-body organism is a kinematic multi-tool Swiss army knife of evolution. 
Its emergent wheel-within is only one of the tools that it can take out or tuck away at will. 
Also at the microscale the wheel-within hides, for instance in
microtubules, where the ZEEM originates in the structure of the protein protofilaments 
making up the microtubule's lattice\cite{MT1,MT2} or in filamentous viruses,
where stresses due to attachment of sugary chains from the mucus
to the virus's spike proteins induce ZEEMs\cite{torovirusSM}.

So when it comes to rivaling man-made technology, 
Nature's "Mount Improbable"\cite{Dawkins} -- in the light of the wheel-within -- appears 
as a misnomer for the epistemological blind-spot caused by our own "Mount Incomprehensible". 
In contrast to the molecular wheel-axle systems readily utilized in Nature 
(see the examples discussed below), the various wheels-within from Fig.~\ref{fig1} 
are kinematically difficult to comprehend at first. Yet, in spite of their intricacy, 
they have reemerged in different clades of life, from 
bacteria, plants, insects to vertebrates
at least four times independently. So the motif forms an evolutionary reachable, 
though not quite low-hanging fruit, as it requires a subtle coordination/orchestration 
of the driving mechanism.

Hardly any example for Nature's current technological superiority, stands out more 
then the three (sic!) distinct types of wheels integrated as parts of the bacterial flagellum 
machinery complex\cite{Namba-1,Bact-Hook,Namba-2}. 
The flagellum is the spot where the classical wheel-axle and the wheel-within 
work in close harmony. In fact, two types
of classical wheel-axle systems, mutually geared together, are coupled with a wheel-within. 
One of these (classic) wheels comes in many copies as an active proton driven rotary engine, 
while the other is a much larger, passive rotor to which the little motorized wheels 
cooperatively couple in order to coordinate and amplify their torque 
contributions\cite{HBerg_review}. 
Right next to them, however, is the third wheel -- the bacterial flagellar hook, a wheel-within --
transmitting the motions of the classical wheels around a $90^\circ$ corner 
in a highly controlled manner to drive the rotation of the propeller unit (the flagellum). 
While the two classical wheels are very reminiscent of man-made technology, 
Nature literally "flexes" with its hook:  
When it comes to the wheel-within, there is nothing even remotely as refined 
as the flagellar hook. While it has been compared 
to an ``ideal'' or ``universal joint"\cite{Namba-1}, 
unlike the hook an ideal joint has no ability to intrinsically maintain its curved geometry. 
In fact, currently we have no hint on how to rebuild the full functionality of the hook's ZEEM 
on the macroscopic scale. Yet, extruding such a self-standing structure with an extended ZEEM along its contour, would be highly beneficial. 
Such a hypothetical infinitely long industrially produced filament 
would form an ideal wheel-within, waiting to be driven in simple geometries, 
via light or other stimuli.
Having a single intrinsic ZEEM 
would allow for much simpler, easier to control drive than 
the 
``ZEEM + modulator mode'' structures 
that have to undergo a dissipative bifurcation first. 
While the closed, toroidal fiberdrive seemingly solves this problem, 
it does so at the price of its own closure limiting its practical flexibility and industrial scalability.     

Another even more fundamental concept from Nature turning around a 
(this time classical) wheel-axle is the conversion of proton gradients to 
the synthesis of ATP in the "ATP Synthase" machinery of our mitochondria. 
The brilliant solution of coupling two fluxes (protons and ATP) 
that are chemically difficult to couple
by utilizing the mechanics of rotation is extraordinarily elegant.
The ZEEMs of a wheel-within have also the potential to couple multiple physico-chemical fluxes 
via their continuously evolving geometry. Elastic chemical engines built with elastomers, 
wheels and pulleys have been created 
in the past\cite{steinberg_mechanochemical_1966,pines_mechanochemical_1973}, 
yet the dream of reproducing a mechanosynthesis similar as achieved by ATP Synthase 
has not reached so far. 
Wet, solvent soaked wheels-within of the future, modifying mechanically the chemical reaction 
rates in cyclic manner could open promising routes to meet the challenge. The race is on.

\section*{References}
\bibliography{bib_for_ZEEM_clean_fused_fin.bib}

%\section*{Code availability}

\section*{Acknowledgements}
~IMK acknowledges funding by ANR-DFG Grant Rodrolls. 
He thanks Herv\'e Mohrbach, Jens-Uwe Sommer and Helmut Schiessel 
and FZ Ulrich S.~Schwarz for many fruitful discussions and constant support.  

\section*{Authors contributions}
~IMK and FZ conceived the ideas and co-wrote the paper.
\section*{Competing interests}
~The authors declare no competing interests.

\end{document}